\documentclass[10pt,conference]{IEEEtran}
\IEEEoverridecommandlockouts
\usepackage{cite}
\usepackage{amsmath,amssymb,amsfonts}
\usepackage{algorithmic}
\usepackage{graphicx}
\usepackage{textcomp}
\usepackage{xcolor}
\usepackage{array}
\usepackage{textcomp}
\usepackage{stfloats}
\usepackage{url}
\usepackage{verbatim}
\usepackage{mathtools}
\usepackage[T1]{fontenc}
\usepackage{booktabs}
\usepackage[nolist]{acronym}
\usepackage {setspace}
\ifCLASSOPTIONcompsoc
\usepackage[caption=false,font=normalsize,labelfon
t=sf,textfont=sf]{subfig}
\else
\usepackage[caption=false,font=footnotesize]{subfi
	g}
\fi

\def\BibTeX{{\rm B\kern-.05em{\sc i\kern-.025em b}\kern-.08em
    T\kern-.1667em\lower.7ex\hbox{E}\kern-.125emX}}

\begin{document}

\title{Modeling and Design of the Communication Sensing and Control Coupled\\ Closed-Loop Industrial System
\vspace{-0.4cm}
}

\author{\IEEEauthorblockN{ Zeyang Meng, Dingyou Ma, Shengfeng Wang, Zhiqing Wei and Zhiyong Feng}
	\IEEEauthorblockA{\textit{Beijing University of Posts and Telecommunications}\\
		\textit{Key Laboratory of the Universal Wireless Communications}, 
		Beijing, China \\
		\{mengzeyang, dingyouma, sfwang, weizhiqing, fengzy\}@bupt.edu.cn}
  \vspace{-1cm}
}

\maketitle

\begin{abstract}
With the advent of 5G era, factories are transitioning towards wireless networks to break free from the limitations of wired networks.
In 5G-enabled factories, unmanned automatic devices such as automated guided vehicles and robotic arms complete production tasks cooperatively through the periodic control loops.
In such loops, the sensing data is generated by sensors,
and transmitted to the control center through uplink wireless communications. 
The corresponding control commands are generated and sent back to the devices through downlink wireless communications.
Since wireless communications, sensing and control are tightly coupled, there are big challenges on the modeling and design of such closed-loop systems.
In particular, existing theoretical tools of these functionalities have different modelings and underlying assumptions, which make it difficult for them to collaborate with each other. Therefore, in this paper, an analytical closed-loop model is proposed, where the performances and resources of communication, sensing and control are deeply related.
To achieve the optimal control performance, a co-design of communication resource allocation and control method is proposed,  inspired by the model predictive control algorithm.
Numerical results are provided to demonstrate the relationships between the resources and control performances.
\end{abstract}

\begin{IEEEkeywords}
control loop, wireless network, effective capacity, estimation theory, model predictive control
\end{IEEEkeywords}

\begin{spacing}{0.96}
\vspace{-0.3cm}
\section{Introduction}
\begin{figure}
	\centering
	\includegraphics[width=0.25\textheight]{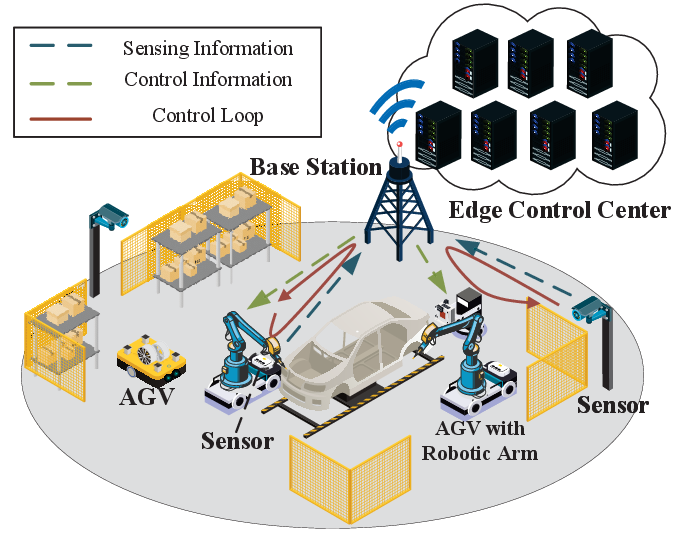}
	\DeclareGraphicsExtensions.
    \vspace{-0.25cm}
	\caption{A 5G enabled factory, with unmanned automatic devices, complete tasks through the collaboration of wireless communication, sensing and control.}	
    \vspace{-0.6cm}
	\label{introductionFig}
\end{figure}

\vspace{-0.2cm}
The emergence of 5G technology is revolutionizing the traditional production modes, offering unprecedented opportunities for the factories to achieve greater productivity and profitability.
As a typical application scenario of 5G, factories are undergoing a transformation from wired networks to wireless networks with the aid of 5G mobile system.
The low latency, exceptional bandwidth and high reliability features of 5G networks enable to reduce the wiring costs, improve the equipment flexibility, and support more highly automated equipment to participate in the production process \cite{5G_Industry_1, 5G_Industry_2}.
An example of the 5G enabled factory is shown in Fig.~\ref{introductionFig}.
In this scenario, unmanned automatic devices such as \acp{AGV} and robotic arms complete production tasks cooperatively, where the control loops are carried out periodically as Fig. \ref{introductionFig}. 
Sensors, located on the devices or settled independently in the factories, collect the states of devices, and forward them to the edge control center. 
Sensing data from multiple sources is fused at the edge to provide more accurate estimation of device status.
The control strategies are then calculated in the control center according to the global sensing data and sent back to the devices \cite{5G_standard_22261}.

The modeling and design of such system face big challenges, since the performances of communication, sensing and control are tightly coupled. 
For example, a large amount of sensing data results in an accurate state estimation in the control center, but causes heavy communication loads in the uplink channel. 
Besides, the uncertain delay and package loss of the wireless channels under limited resources greatly affect the stability and convergence of control. 
Furthermore, the uplink and downlink communications of such closed-loop system are closely related through the generated control strategies. 
Therefore, a joint design of such close-loop control system is required, rather than designing communication, sensing, and control in a separate manner.
The design and modeling of such closed-loop control systems are investigated in the works of networked control systems \cite{NCS1,NCS2,Leveraging,NCSAdd1, NCSAdd2, NCSAdd3}  and unmanned aerial vehicle networks \cite{UAV1}. 
However, in these studies, the state information of the devices is usually assumed to be perfectly sensed. Besides, the communication in such systems are usually regarded as independent tunnels, which ignores the potential tradeoffs between communication, sensing and control.

In this paper, a joint sensing, communication and control model of the closed-loop system is proposed.
The closed-loop performances such as cycle time and package loss probability are derived with respect to the physical layer resources.
In addition, a stability-communication-sensing inequality is derived to reveal the communication, sensing and control requirements for system convergence.
%
To achieve the optimal control performance, the resource allocation of communication and the control method are jointly designed. 
Numerical results are utilized to verify the effectiveness of the proposed method,  and to demonstrate the relationship between the resource allocation and control performance.

The rest of the paper is organized as follows. Section~\uppercase\expandafter{\romannumeral2} presents the system model of the closed-loop system. In Section~\uppercase\expandafter{\romannumeral3}, the joint design of the resource allocation and control method is proposed. Numerical simulations are utilized to evaluate the proposed system and method in Section~\uppercase\expandafter{\romannumeral4}.  Finally, Section~\uppercase\expandafter{\romannumeral5} provides concluding remarks.

\begin{figure}
	\centering
	\includegraphics[width=0.3\textheight]{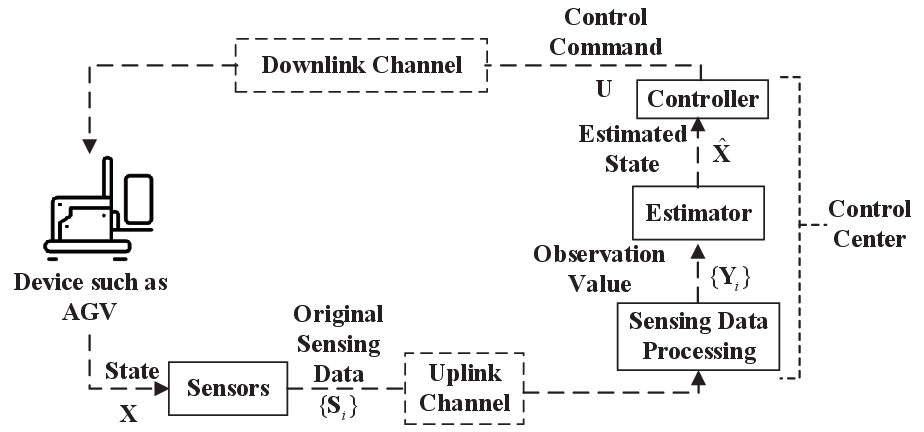}
	\DeclareGraphicsExtensions.
    \vspace{-0.25cm}
	\caption{System Model}	
    \vspace{-0.6cm}
	\label{systemModel}
\end{figure}

\vspace{-0.15cm}

\section{System Model}
\label{Sec2}
\vspace{-0.15cm}
A model of the control loop,  as  shown in Fig. \ref{systemModel}, is constructed by referring to the overall model of the networked control system \cite{NCS2}.
The closed-loop control system, which is composed of the automated devices, the sensors, the control center, and two communication links, aims at generating appropriate commands based on the device state sensed by the sensors, and realize on-demand control of the device. 
The loop starts from the automated device which can be an \ac{AGV} or a robotic arm. 
The state information, such as the location and speed, is measured by the sensors mounted in the factory. The original sensing data is transferred to the control center located at the edge by wireless channels.
The control center is a high-performance server, which is utilized to process the sensing data, estimate the state of device, and generate the control commands. After transmitted to the control center, the sensing data is processed in the control center to generate the  observation values of the device state from each sensor. The observation values are then merged to obtain the final state estimation of the device. 
According to the estimated state, the control commands are generated, and the communication resources are allocated to the uplink and downlink at the controller. The control commands generated by the control center are next sent to the devices through the downlink channel to complete the control loop. 
The closed-loop process is repeated continuously to ensure that the devices react quickly to the environmental changes. 

In the sequel, the single models and performance metrics of the sensing, communication, and control are introduced, respectively.

\subsection{Sensing Model}
The purpose of sensing stage is to collect the device's states with the on-board and off-board sensors, so as to provide prior information for decision-making in the control center.
Considering the constraints such as power consumption, weight, and cost, sensors cannot possess strong computing capabilities. 
Therefore, the raw sensing data is  first transmitted to the control center for further processing.
Subsequently, the control center calculates and integrates the global sensing data to estimate the state of the device.

Based on the above process, the signal model of the sensing stage is presented as follows. As Fig.~\ref{systemModel}, the current state $ \mathbf{X}$ of the device is assumed to be observed by $k_s$ sensors per loop, where the original sensing data is denoted by $\{\mathbf{S}_1, \mathbf{S}_2, \cdots , \mathbf{S}_{k_s}\}$.
After processing at the control center, $k_s$ observation values $\{\mathbf{Y}_1, \mathbf{Y}_2, \cdots , \mathbf{Y}_{k_s}\}$ of the state are generated, which satisfies
\vspace{-0.2cm}
\begin{equation}
	\mathbf{Y}_{i}= \mathbf{X} + \mathbf{N}_{i} \quad(1\le i \le k_s),
    \vspace{-0.2cm}
\end{equation}
where $\mathbf{N}_{i}$ is the Gaussian white noise of observations. Without loss of generality, the noises $\left\{\mathbf{N}_i\right\}$ are assumed to have the same distribution and with the same variance $\sigma^2$. The maximum likelihood estimation is utilized to estimate the state of the automated device using the observation values from different sensors. With the above assumptions, the estimate of $\mathbf{X}$, denoted by $\hat{\mathbf{X}}$, is given by 
\cite{StochasticSigPro}
\vspace{-0.2cm}
\begin{equation}
	\hat{\mathbf{X}} = \frac{1}{k_s} \sum_{i=1}^{k_s} \mathbf{Y}_i.
    \vspace{-0.2cm}
\end{equation}
The mean-square error of $\hat{\mathbf{X}}$ is expressed as
\vspace{-0.2cm}
\begin{equation}
	\label{Var}
	\mathbb{E} \{(\hat{\mathbf{X}}-\mathbf{X})^2\} = \frac{1}{k_s}\sigma^2.
    \vspace{-0.2cm}
\end{equation}

\subsection{Communication Model}
In a closed-loop system, communication plays a pivotal role in the transmission of sensing data from the device to the control center, and distributing the  control commands from the control center to the device, as shown in the uplink and downlink of Fig.~\ref{systemModel}.
The uplink and downlink communications are intimately coupled within the control loop, since the control commands transmitted by the downlink channels are generated based on the estimated state of the devices using the sensing data conveyed by the uplink channels. 
Therefore, a closed-loop communication model is established and studied in this section. The communication process is modeled as two queues in series, corresponding to the uplink and downlink communications, respectively.

The arrival data of the uplink queue is originated from the sensing information. For ease of analysis, the sensing data is assumed to be of the same size, which is denoted by $\beta$ bits.
Assume that the device is repeatedly sensed every $T_s$ seconds. The sensing system generates $\frac{\beta k_s}{T_s}$ bits of data per second, which means that the arrival rate of the uplink queue is 
\vspace{-0.2cm}
\begin{equation}\label{Arru}
	\lambda_u =\frac{\beta k_s}{T_s}.
	\vspace{-0.2cm}
\end{equation} 

The arrival data of the downlink queue is the control commands calculated from the uplink transmitted data. Therefore, the arrival rate of the downlink queue, denoted by $\lambda_d$, is the amount of data departing from the uplink queue per second, i.e. the uplink departure rate $L_u$, multiplied by a coefficient, which is 
\vspace{-0.2cm}
\begin{equation}\label{Arrd}
	\lambda_{d} = \frac{N\gamma}{\beta k_s} L_u ,
 \vspace{-0.2cm}
\end{equation}
where the coefficient $\frac{N\gamma}{\beta k_s}$ denotes that every $N$ control commands are generated from $k_s$ sensing data, with $\gamma$ being the size of each control package, and $N$ being the number of control commands generated from the control center.

The service rate of the queue denotes the average number of items that can be served by the queue per unit of time.
In the proposed communication model, the service rate is the average number of packages transmitted per second, i.e. the transmission rate of the communication channel.
Consider that the sensing and control packages transmitted in factories are mainly short packets \cite{IoTURLLC}, the capacity under finite blocklength is applied to be the service rate $\{R_i|i=u,d\}$ of the uplink and downlink queues, i.e. \cite{ShortPacket}
\vspace{-0.2cm}
\begin{equation}
	R_i = W_i \log _2\left(1+\mathrm{SNR_i}\right)-\sqrt{\frac{V_i}{L_i}} f_{\mathrm{Q}}^{-1}\left(e_i\right) \ (i=u,d), 
\vspace{-0.1cm}
\end{equation}
where $W_i$ is the bandwidth of the communication channel, $L_i$ is the blocklength of the packages, $e_i$ is the probability of the irreparable distortion occurs, $f_{\mathrm{Q}}^{-1}$ is the inverse of complementary Gaussian cumulative distribution function, $V_i$ is the channel dispersion, which can be approximated by
\vspace{-0.2cm}
\begin{equation}
	V_i=1-\frac{1}{\left(1+\mathrm{SNR_i}\right)^2} \ \left(i=u,d\right).
\vspace{-0.2cm}
\end{equation}

We next discuss the metrics to evaluate the performance of the uplink and downlink communictions.

Based on the above communication model, closed-loop performance metrics such as the cycle time and the package loss probability are derived later to describe the intricate feedback mechanism of the closed-loop system.
The cycle time is the time for a system to complete one control loop as Fig. \ref{systemModel} \cite{NIST}, which reveals the system's responsiveness to unexpected events.
The package loss probability stands for the probability that the communication is completed exceeding the required time threshold, on the premise that the packages are dropped in an overtime transmission, which affects the control efficiency of the system.

To reveal the relationship between two metrics and physical-layer parameters, the effective capacity theory is applied in the following analysis, where the effective capacity is defined to be the maximum acceptable arrival rate of the communication queues given by \cite{EffectiveCapacity}
\vspace{-0.1cm}
\begin{equation}
	\begin{small}
		\begin{aligned}
			C_i(\theta_i, W_i, e_i) = &-\frac{1}{\theta_i}\log(\mathbb{E}\{\exp(-\theta_i R_i)\}) \quad (i=u,d),
		\end{aligned} 
	\end{small}
    \vspace{-0.1cm}
\end{equation}
where $C_u$ and $C_d$ represent the effective capacities of the uplink and downlink communications, respectively, $\theta_i$ represents the decay rate of queue overflow probability, which satisfies
\begin{equation}
	\lim _{q_0 \rightarrow \infty} \frac{\ln \operatorname{Pr}\{q\left(\infty\right) \geq q_0\}}{q_0} = -\theta_i \quad \left(i=u,d\right) ,
\end{equation}
with $q\left( \infty \right)$ being the length of the communication buffer queue in steady state, and $q_0$ being the buffer overflow threshold \cite{EffectiveCapacity}.
Both $\theta_u$ and $\theta_d$ are assumed to be constant in the industry scenario \cite{IIoTEffectiveCapacity}.

According to the effective capacity theory, the cycle time and package loss probability are derived as Theorem \ref{theoremCom}.

\newtheorem{theorem}{Theorem}
\begin{theorem}[Cycle Time and Package Loss Probability] \label{theoremCom}
	The maximum value of the cycle time $D_{c, \max}$ and package loss probability $\epsilon_c$ of the control loop can be approximated as
	\begin{align}
			D_{c, \max}&=D_{\mathrm{u}, \max }+D_{\mathrm{d}, \max },\\
		\epsilon_c &\approx 1-\left(1-\epsilon_{u}\right)\left(1-\epsilon_{d}\right), 
        \vspace{-0.1cm}
    \end{align}
	where $\epsilon_u$ and $\epsilon_d$ are the package loss probabilities of uplink and downlink queues, 
	$D_{\mathrm{u}, \max}$ and $D_{\mathrm{d}, \max}$ are the maximum delays of uplink and downlink queues,
	and
	$D_{\mathrm{u}, \max}$, $D_{\mathrm{d}, \max}$, $\epsilon_u$, and $\epsilon_d$ satisfy
   \vspace{-0.15cm}
	\begin{equation}
		D_{i, \max }=-\frac{\ln \left(\epsilon_i\right)}{\theta_i C_i\left(\theta_i, W_i, e_i\right)}\quad (i=u,d) .
  \vspace{-0.15cm}
	\end{equation}
\end{theorem}
\begin{IEEEproof}
	The proof is provided in  Appendix A.
\end{IEEEproof}

Besides, Theorem \ref{theoremEC} holds according to the definition of the effective capacity.
\begin{theorem}[Inequality of Effective Capacity] \label{theoremEC}
	According to the definition of the effective capacity, $C_u\left(\theta_u, W_u, e_u\right)$ and $C_d\left(\theta_d, W_d, e_d\right)$ should satisfy
    \vspace{-0.1cm}
	\begin{equation}\label{IneqCu}
		\begin{aligned}
			C_u\left(\theta_u, W_u, e_u\right) \ge  \lambda_{u}(\beta,k_s,T_s)
			=  \frac{\beta k_s}{T_s},
		\end{aligned}
        \vspace{-0.1cm}
	\end{equation}
	and 
    \vspace{-0.1cm}
	\begin{equation}\label{IneqCd}
		\begin{aligned}
			&C_d\left(\theta_d, W_d, e_d\right) \ge \lambda_{d}(\beta,k_s,T_s,\theta_u,\theta_d,W_u,e_u)\\
			&\qquad= \left\{\begin{array}{l}
				\frac{N\gamma}{\beta k_s}\lambda_{u}(\beta,k_s,T_s), \quad 0 \leq \theta_d \leq \theta_u \\
				\frac{N\gamma}{\beta k_s\theta_d}\{\left(\theta_d-\theta_u\right) C_{u}\left(\theta_u-\theta_d, W_u, e_u\right)\\
				+\lambda_{u}(\beta,k_s,T_s) \theta_u\}, \quad \theta_d>\theta_u
			\end{array}\right..
		\end{aligned}
        \vspace{-0.1cm}
	\end{equation}
	
\end{theorem}
\begin{IEEEproof}
	The proof is given in Appendix B.
\end{IEEEproof}

\vspace{-0.2cm}
\subsection{Control Model}
\vspace{-0.1cm}

The purpose of the control stage is to generate the control commands based on the current state of the device, so as to enable the device to reach the expected state after a period of time.
In the control field, the state functions are often applied to model the state evolution.  
The following subsections progressively establish the state function of \ac{AGV}, 
advancing from the ideal model to the model with imperfect sensing, and finally to the model influenced by the imperfect wireless communication.

\subsubsection{Ideal Model of \ac{AGV}}
Denote the state of the \ac{AGV} as $\mathbf{X}=\left[\delta, v, a\right]^T$, with $\delta$, $v$, and $a$ being the position, velocity, and acceleration, respectively.
The controller, which is denoted by $\mathbf{U}$, is designed to be the linear combination of the state $\mathbf{X}$, i.e. $\mathbf{U} = \mathbf{K}\mathbf{X}$, where $\mathbf{K} = \left[K_{1}, K_{2}, K_{3}\right]$ denotes the coefficient matrix of the linear transformation.
The state function of the \ac{AGV} is expressed as follows  \cite{ControlModel}
\begin{equation}
	\label{continousControlMocel}
	\dot{\mathbf{X}}=\mathbf{A} \mathbf{X}+\mathbf{B} \mathbf{U}, 
    \vspace{-0.1cm}
\end{equation}
where $\dot{\mathbf{X}}$ denotes the differential of the state $\mathbf{X}$, $\mathbf{A}$ represents the state matrix, and $\mathbf{B}$ is the control matrix. The expressions of $\mathbf{A}$ and $\mathbf{B}$ are given by 
\begin{equation}
	\mathbf{A}=\left[\begin{array}{ccc}0 & 1 & 0 \\ 0 & 0 & 1 \\ 0 & 0 & -1 / \varsigma\end{array}\right],
\end{equation}
and 
\begin{equation}
	\mathbf{B}=\left[\begin{array}{c}0 , 0 , -1 / \varsigma\end{array}\right]^\mathrm{T},
\end{equation}
respectively, where $\varsigma$ being a constant related to the engine. In \eqref{continousControlMocel}, $\mathbf{A}\mathbf{X}$ denotes the evolution of the device states without control involved, and $\mathbf{B}\mathbf{U}$ denotes the change of states owing to the control commands.

For ease of analysis, \eqref{continousControlMocel} is discretized with the Euler's method \cite{Euler} as  
\vspace{-0.1cm}
\begin{equation}
	\begin{aligned}\label{discretizedOrigControlFunc}
		\frac{1}{T_d}(\mathbf{X}_{t+1}-\mathbf{X}_{t}) = \mathbf{A} \mathbf{X}_{t}+\mathbf{B} \mathbf{U}_{t}\\
	\end{aligned},
    \vspace{-0.1cm}
\end{equation}
by substituting $\dot{\mathbf{X}}$ with $\frac{1}{T_d}(\mathbf{X}_{t+1}-\mathbf{X}_{t})$, where $\mathbf{X}_{t}$ is the state at time $t$, $\mathbf{U}_{t}$ is the control command at time $t$, and $T_d$ is the time step where the state of the device is assumed to be constant.
\eqref{discretizedOrigControlFunc} can be reorganized as
\begin{equation}\label{reorgDisCtrlFunc}
	\begin{aligned}
		\mathbf{X}_{t+1} = \widetilde{\mathbf{A}}\mathbf{X}_{t}+\widetilde{\mathbf{B}} \mathbf{U}_{t} ,
	\end{aligned}
    \vspace{-0.1cm}
\end{equation}
where $\widetilde{\mathbf{A}}=T_d \mathbf{A}+\mathbf{I}$, and $\widetilde{\mathbf{B}}=T_d\mathbf{B}$.

\addtolength{\topmargin}{0.02in}
\subsubsection{The Model Considering Imperfect Sensing}
The imperfect estimation makes it difficult for the controller that generates the control commands based on the accurate perceptual information, thus results in the control deviation.
Taking into account the effect of the imperfect estimation at the control center, the device receives the estimated-state-based control command $\hat{\mathbf{U}}_t = \mathbf{K}\hat{\mathbf{X}}_t$ rather than $\mathbf{U}_t$.
Therefore, \eqref{reorgDisCtrlFunc} can be developed as 
\vspace{-0.1cm}
\begin{equation}\label{sensingEffectCtrlFunc}
	\begin{aligned}
		\mathbf{X}_{t+1} = \widetilde{\mathbf{A}}\mathbf{X}_{t}+\widetilde{\mathbf{B}} \hat{\mathbf{U}}_{t}.
	\end{aligned}
    \vspace{-0.1cm}
\end{equation}

\subsubsection{The Model Considering Imperfect Sensing And Wireless Communication}
The effect of wireless communication can be attributed to the inaccurate control caused by communication delay and the loss of control instructions due to the package loss.

When communication delay occurs, the device will receive control commands corresponding to the previous state rather than the current state, resulting in a suboptimal control strategy.
However, the delay-compensated strategy can be applied to compensate for the impact of communication delay \cite{DelayedMPC}.
Briefly speaking, the control center not only calculates control commands based on the received sensing data but also predicts the future states and corresponding control commands for the subsequent time steps utilizing the state function and transmits them collectively to the device.
The device calculates the cycle time based on the time difference between sensing and receiving control commands, subsequently selecting and executing the control command corresponding to the cycle time.

Although the effect of the communication can be compensated with the above algorithm, the effect of package loss in wireless transmission still remains non-negligible.
When the package loss occurs, the control commands will not be received by the device, then the state function \eqref{sensingEffectCtrlFunc} is revised as
\vspace{-0.1cm}
\begin{equation}
	\begin{aligned}\label{stateFunc_complex}
		\mathbf{X}_{t+1} = \widetilde{\mathbf{A}}\mathbf{X}_{t}+\eta\widetilde{\mathbf{B}} \hat{\mathbf{U}}_{t} ,
	\end{aligned}
    \vspace{-0.1cm}
\end{equation}
where $\eta$ is a factor related to the packet loss that equals to $1$ with probability $1-\epsilon_c$ and $0$ with probability $\epsilon_c$.

We next discuss the performance metric of the control system. 
A feasible control system is needed to be convergence, i.e. the state of the controlled devices reaches the target state within a certain period of time.
Therefore, under the proposed delay-compensated strategy and discretized state function, the following theorem is proposed based on \eqref{Var}, \eqref{stateFunc_complex}, and Lyapunov stability theory \cite{LyaTheory} to guarantee the asymptotic convergence under communication delay, package loss, and sensing error.
\begin{theorem}[Stability-Communication-Sensing Inequality]\label{theoremCtrl}
	The system is asymptotic convergent when $\mathbf{K}_t$, $\epsilon_c$ and $\sigma$ satisfies
	\begin{equation}  \label{jointFunction}
		(1-\epsilon_c)  F_1(\mathbf{X}_t, \mathbf{K}_t,\sigma) \leq F_2(\mathbf{X}_t),
	\end{equation}
	where
 \vspace{-0.1cm}
	\begin{equation}
		\begin{small}
			\begin{aligned}
				\!\!\!\!\!F_1&(\mathbf{X}_t, \mathbf{K}_t, \sigma) = \left(\widetilde{\mathbf{A}} \mathbf{X}_t+\widetilde{\mathbf{B}} \mathbf{K}_t \mathbf{X}_{t}\right)^\mathrm{T} \cdot\mathbf{P}\left(\widetilde{\mathbf{A}} \mathbf{X}_t+\widetilde{\mathbf{B}} \mathbf{K}_t \mathbf{X}_{t}\right)\\
				&\!\!\!\!\!+\operatorname{Tr}\left[(\widetilde{\mathbf{B}} \mathbf{K}_t)^\mathrm{T} \mathbf{P}(\widetilde{\mathbf{B}} \mathbf{K}_t) \cdot \frac{1}{k_s}\sigma^2\right]
				-\left(\widetilde{\mathbf{A}} \mathbf{X}_t\right)^\mathrm{T} 	\mathbf{P}\left(\widetilde{\mathbf{A}} \mathbf{X}_t\right),
			\end{aligned}
		\end{small}
	\vspace{-0.15cm}
	\end{equation}
\vspace{-0.1cm}
	\begin{equation}
		\begin{small}
			\begin{aligned}
				F_2(\mathbf{X}_t) &=  \mathbf{X}_t^\mathrm{T} \mathbf{P} \mathbf{X}_t-\left(\widetilde{\mathbf{A}} \mathbf{X}_t\right)^\mathrm{T} 	\mathbf{P}\left(\widetilde{\mathbf{A}} \mathbf{X}_t\right),
			\end{aligned}
		\end{small}
	\vspace{-0.1cm}
	\end{equation}	
\end{theorem}
\vspace{-0.1cm}
with $\mathbf{P}$ being the constant positive semidefinite matrices, whose value is determined based on the actual system.
\begin{IEEEproof}
	See the proof in Appendix C.
\end{IEEEproof}

\addtolength{\topmargin}{0.05in}

\section{Joint Design of Resource Allocation and Control Method}
\label{SecIII}
According to the above analysis, the performances of communication, sensing and control are tightly united. 
The system needs to be globally optimized to achieve the closed-loop optimization.

Therefore, to achieve the optimal control performance under the impact of limited resources, imprecise sensing, and package losses, a model-predictive-control-based joint design of the resource allocation and control method is carried out at the control center,
aiming at guaranteeing the minimum control cost.
The control cost is given as follows in the form of a quadratic function \cite{MPCBook}
\vspace{-0.2cm}
\begin{equation}
	J(\mathbf{X}_t,\mathbf{U}_t) = \sum_{t=1}^{N-1}\left(\mathbf{X}_t^\mathrm{T} \mathbf{P} \mathbf{X}_t+\mathbf{U}_t^\mathrm{T} \mathbf{R} \mathbf{U}_t\right)+\mathbf{X}_N^\mathrm{T} \mathbf{S} \mathbf{X}_N,
    \vspace{-0.2cm}
\end{equation}
with $N$ being the time horizon over which the control actions are optimized.
The term $\mathbf{U}_t^\mathrm{T} \mathbf{R} \mathbf{U}_t$ represents the energy required in the control actions at time $t$.
The term $\mathbf{X}_t^\mathrm{T} \mathbf{P} \mathbf{X}_t$ and $\mathbf{X}_N^\mathrm{T} \mathbf{S} \mathbf{X}_N$, which can be reformed as $(\mathbf{X}_t-\mathbf{0})^\mathrm{T} \mathbf{P} (\mathbf{X}_t-\mathbf{0})$ and $(\mathbf{X}_N-\mathbf{0})^\mathrm{T} \mathbf{S} (\mathbf{X}_N-\mathbf{0})$, are the distances to the control objective $\mathbf{X}=\mathbf{0}$ at time $t$ and $N$, respectively.
$\mathbf{R}$ and $\mathbf{S}$ represent the weights of the above three terms. 
The choice of $\mathbf{P}$, $\mathbf{R}$, and $\mathbf{S}$ depends on the system dynamics and the design objectives, which are often designed to be positive semidefinite to ensure convexity \cite{LQRDesign}.

The joint design of the resource allocation and control method can be expressed as the following optimization problem given by
\vspace{-0.15cm}
\begin{subequations}\small\label{P1}
	\begin{alignat}{2} 
		\text{P1:} &\min\limits_{\substack{\mathbf{K}_t,W_i,\epsilon_i \\ (i=u,d)}} \quad && J(\mathbf{X}_t,\mathbf{U}_t)  \label{P1a}\\ 	
		& \quad \ \text { s.t. } \quad&& C_u \geq \lambda_u(\beta,k_s, T_s) \label{P1b}\\
		&&& C_d \geq \lambda_d(\beta,k_s, T_s, \theta_u,\theta_d,W_u,e_u) \label{P1c}\\
		&&& D_{c,\mathrm{max}} \le D_0\label{P1d}\\
		&&& (1-\epsilon_c)  F_1(\mathbf{X}_t, \mathbf{K}_t, \sigma) \leq F_2(\mathbf{X}_t) \label{P1e}\\
		&&& 0 \le W_u + W_d \leq W_0 \label{P1f}\\
		&&& \mathbf{U}_t = \mathbf{K}_t \mathbf{X}_t \label{P1g}\\
		&&& 0\le\epsilon_c\le1 \label{P1h}\\
		&&& \mathbf{X}_{t+1} = \widetilde{\mathbf{A}} \mathbf{X}_t + (1-\epsilon_c)\widetilde{\mathbf{B}} \mathbf{U}_t \label{P1i}\\
		&&& \mathbf{X}_1 = \mathbf{X}_{\text{ini}} \label{P1j},
	\end{alignat}
\end{subequations}
where $W_0$ and $D_0$ are thresholds of the total bandwidth and cycle time.
\eqref{P1b} and \eqref{P1c} ensure that the effective capacities of the uplink and downlink channels are greater than the arrival rates, which are stated at Theorem \ref{theoremCom}. 
\eqref{P1d} ensures that the cycle time of the control loop is constrained in $D_{0}$.
\eqref{P1e} is the constraint on the package loss probability, sensing error, and convergence, which is derived in Theorem \ref{theoremCtrl}.
\eqref{P1f} constrains the available bandwidth of the communication.
\eqref{P1g} is the relationship between the control commands and the state, which is stated in the control model.
\eqref{P1i} is the exception of the state function of the device, which is utilized to predict the future state.
\eqref{P1j} is the initial condition of the state $\mathbf{X}_t$, with $\mathbf{X}_{\text{ini}}$ being a constant matrix.

Problems of such objective function with decision variables $\mathbf{K}_t$ is proved to be a non-convex nonlinear programming problem.
Typical methods for solving such problems involve employing gradient projection or interior point algorithms to find the suboptimal points \cite{NonCov}.
In the following section, the interior point algorithm is applied to solve the above problem.

\vspace{-0.12cm}
\section{Numerical Results}
\begin{figure*}[!t]
	\centering
	\subfloat[Position trajectories of \ac{AGV} with different numbers of sensing samples $k_s$]{\includegraphics[width=2in]{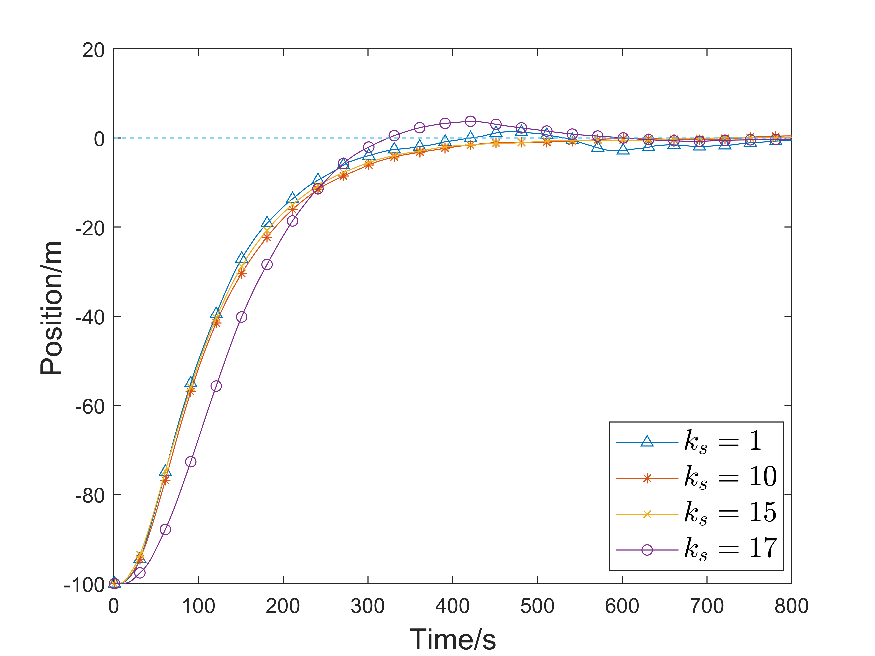}
		\label{Sim_ks}}
	\hfil
	\subfloat[Position trajectories of \ac{AGV} with different available $W_0$]{\includegraphics[width=2in]{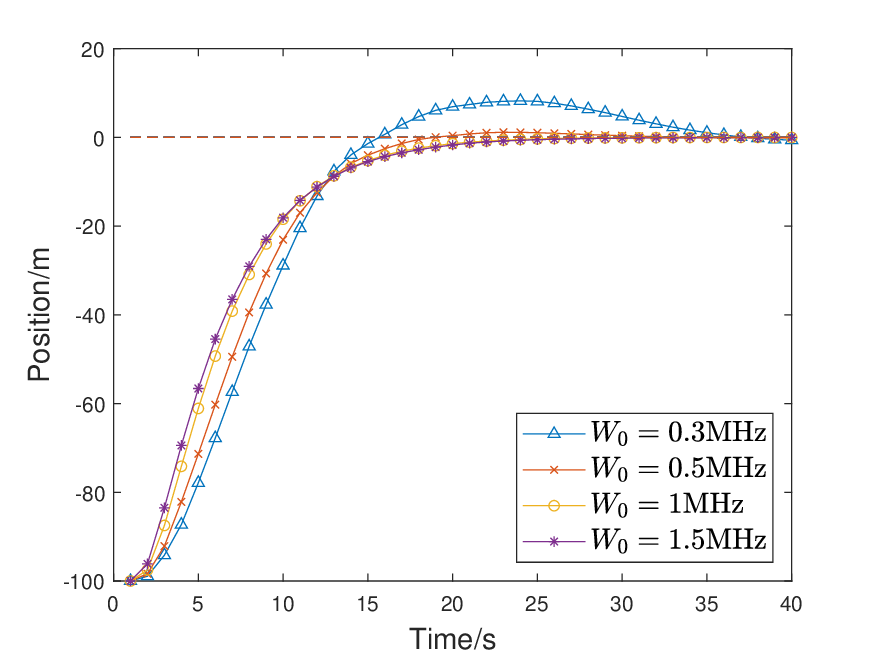}
		\label{Sim_B0}}
	\hfil
	\subfloat[Position trajectories of \ac{AGV} with different delay threshold $D_0$]{\includegraphics[width=2in]{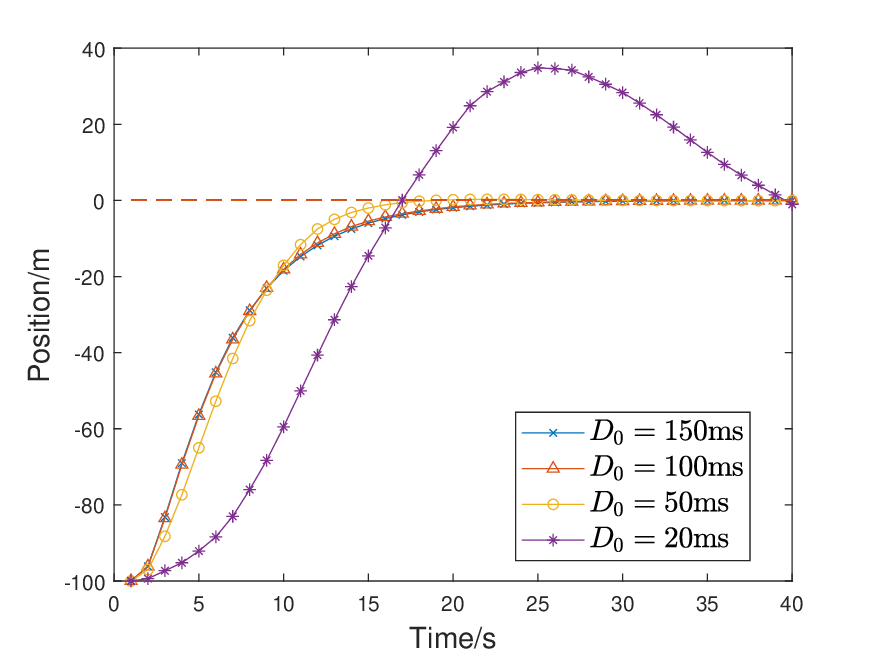}
		\label{Sim_D0}}
	\caption{Numerical results of \ac{AGV} with different parameters}
	\label{BigFig}
    \vspace{-0.6cm}
\end{figure*}

\vspace{-0.1cm}
In this section, the optimizer IPOPT \cite{IPOPT} is applied to obtain the numerical results with interior point algorithm.
Take the brake control of the \ac{AGV} as an example, which needs to stop the \ac{AGV} at the target position $0$.
The resource allocation and controller design are simulated and analyzed in this section. The \ac{AGV} with an initial speed of 0 m/s is required to move forward and stop at the position 100 m ahead, i.e., $\mathbf{X}_{\text{ini}} = [100,0,0]$. The rest of the simulation parameters are presented in TABLE \ref{tableSimPara}, and the trajectories under different parameters are shown in Fig. \ref{BigFig}.
\vspace{-0.4cm}
\begin{table}[h]
	\centering 
	\caption{Simulation Parameters}
    \vspace{=-0.1cm}
	\renewcommand{\arraystretch}{1.1}
    \footnotesize
	\begin{tabular}{m{1.3cm}<{\centering} m{2cm}<{\centering} m{1.3cm}<{\centering} m{2cm}<{\centering}}	
		\toprule[0.5pt]\toprule[0.5pt]
		\textbf{Parameters} & \textbf{Value} & \textbf{Parameters} & \textbf{Value}\\
		\midrule[0.5pt]
		$\sigma$ & 1.5 & $L$ & 200\\
		$\beta$ & 1000 bit & $T_d$ & 0.05 s \\
		$\gamma$ & 20 bit & $T_s$ & 0.05 s\\
		$\mathrm{SNR}_u$ & 20& $\theta_u$ & 0.001\\
		$\mathrm{SNR}_d$ & 30& $\theta_d$ & 0.002\\
		$e_u$ & 0.001 & $\varsigma$ & 0.125\\
		$e_d$ & 0.002 & $N$ & 10 \\
		\bottomrule[0.5pt]\bottomrule[0.5pt]
	\end{tabular}
	\label{tableSimPara}
\end{table}
\vspace{-0cm}

Fig. \ref{BigFig}(a) illustrates the trajectories of \ac{AGV} with different numbers of sensing samples $k_s$, with $W_0 = 1.5$ MHz and $D_{0}=150$ ms.
When $k_s$ is small, there is significant jitter in the state trajectory, especially in the trajectory around $0$, because the small amount of sensing data cannot reduce the effect of the noise of sensing.
When $k_s$ becomes larger, the state trajectories tends to be a smoother profile, and the increase of $k_s$ leads to a faster convergence. 
However, when $k_s$ reaches $17$, the convergence speed of the system decreases sharply, and the trajectory exceeds the target position, which will bring great collision risk when obstacles such as walls are located at the target position.
The reason for this phenomenon is that when $k_s$ is small, the system has sufficient communication resources to complete the uplink sensing services.
In contrast, the increasing number of the sensing samples enhances the accuracy of control instructions and thus improves the speed of state convergence.
When $k_s$ is large, the packet loss rate increases sharply due to the great pressure of the uplink communication and the limited communication resources.
Since more time are consumed in the uplink communications, the control commands are difficult to be transmitted to the device timely.
The benefits on sensing accuracy cannot offset a large number of package losses, resulting in the control disorder shown in Fig. \ref{Sim_ks}.

Fig. \ref{BigFig}(b) illustrates the position trajectories of \ac{AGV} with different available $W_0$, with $k_s = 10$ and $D_{0}=150$ ms.
When the bandwidth resource is sufficient, such as $W_0=1.5$ MHz and $W_0=1$ MHz, sensing and control packages can be transmitted quickly with few packet losses, thus the control performance of the system is basically the same.
When there is no sufficient bandwidth, the packet delay increases significantly, and the packages drop easily, resulting in large trajectory fluctuations and making it converge slowly.

Furthermore, the trajectories of \ac{AGV} with different thresholds of the cycle time $D_0$ are shown in Fig. \ref{BigFig}(c), with $k_s = 10$ and $W_0 = 1.5$ MHz.
The trajectories are nearly the same with different delays unless the delay bound is too harsh to maintain a low package loss rate, which indicates that the delay compensation method adopted in this paper is effective.

\vspace{-0.12cm}
\section{Conclusion}
\vspace{-0.12cm}
In this paper, an analytical model of the closed-loop system with communication, sensing and control coupled was proposed. 
Specifically, the sensing error was measured by the mean-square error of the maximum likelihood estimation.
The communication delay and package loss probability were derived through the effective capacity theory, which relates link layer performances with communication resources.
The control model under communication delay, package loss and sensing error was established afterwards, and the stability-communication-sensing inequality was derived.
Besides, an optimization problem was proposed to allocate the resources and obtain the control commands simultaneously.
The numerical results demonstrated the effectiveness of the proposed algorithm and the complex relationships between system performance and resources.

\vspace{-0.2cm}
\section*{Appendix A \\ Proof of Theorem 1}
\vspace{-0.1cm}

According to \cite{EffectiveCapacity}, the probability that the communication delay $\{D_i | i= d,u\}$ exceeds the delay threshold $D_{i,\max}$ satisfies
\vspace{-0.15cm}
\begin{equation}\label{ProbLoss}
	\begin{small}
		\operatorname{Pr}\left\{D_i \geq D_{i, \max }\right\} \approx e^{-\theta_i C_i\left(\theta_i, W_i\right) D_{i, \max }}=\epsilon_i ,
	\end{small}
 \vspace{-0.15cm}
\end{equation}
where $\epsilon_u$ and $\epsilon_d$ are the probabilities that the communication delays of uplink and downlink communication exceeds $D_u$ and $D_d$ respectively.

According to \eqref{ProbLoss}, the uplink and downlink delays are derived as 
\vspace{-0.15cm}
\begin{equation}
	D_{i, \max }=-\frac{\ln \left(\epsilon_i\right)}{\theta_i C_i\left(\theta_i, W_i, e_i\right)}\quad (i=u,d)
	\vspace{-0.15cm}
\end{equation}

Since the calculation delay of the control commands and the sensing delay of the sensors are relatively small in the control loop, the maximum value of the cycle time $D_c$ can be approximated as the sum of the maximum uplink delay and downlink delay, i.e., 
\vspace{-0.15cm}
\begin{equation}
	D_{c, \max}=D_{u, \max }+D_{d, \max }.
    \vspace{-0.15cm}
\end{equation}

If the delays of the packages exceed the maximum threshold settled, the packages will be dropped. In this case, the package drop probability of the system is 
\vspace{-0.15cm}
\begin{equation}
	\begin{aligned}
		\epsilon_c = 1-\left(1-\epsilon_{u}\right)\left(1-\epsilon_{d}\right) .
	\end{aligned}
    \vspace{-0.15cm}
\end{equation}

\vspace{-0.2cm}
\section*{Appendix B \\ Proof of Theorem 2}
\vspace{-0.1cm}

The inequality \eqref{IneqCu} is derived by the definition that the effective capacity of the uplink communication is greater than the corresponding arrival rate \eqref{Arru}.

As for inequality \eqref{IneqCd}, since the departure process $L_u$ of the uplink queue is \cite[Eqn. 10]{TandemQ}
\vspace{-0.15cm}
\begin{equation} \small \label{DepRateUp}
	\begin{aligned}
		L_u\left(\theta_u, \theta_d\right)=\left\{\begin{array}{l}
			\lambda_{u}, \quad 0 \leq \theta_d \leq \theta_u \\
			\frac{1}{\theta_d}\{\left(\theta_d-\theta_u\right) C_{u}\left(\theta_u-\theta_d, B_u, e_u\right)\\
			+\lambda_{u} \theta_u\}, \quad \theta_d>\theta_u
		\end{array}\right. .
	\end{aligned}
	\vspace{-0.15cm}
\end{equation}
By substituting \eqref{DepRateUp} into \eqref{Arrd}, the arrival rate of the downlink queue $\lambda_{d}$ is given as follows
\vspace{-0.15cm}
\begin{equation} \small
	\begin{aligned}
		\lambda_{d}(\beta,k_s,&T_s,\theta_u,\theta_d,B_u,e_u)= \\
		&\left\{\begin{array}{l}
			\frac{N\gamma}{\beta k_s}\lambda_{u}, \quad 0 \leq \theta_d \leq \theta_u \\
			\frac{N\gamma}{\beta k_s\theta_d}\{\left(\theta_d-\theta_u\right) C_{u}\left(\theta_u-\theta_d, B_u, e_u\right)\\
			+\lambda_{u} \theta_u\}, \quad \theta_d>\theta_u
		\end{array}\right. .
	\end{aligned}
\vspace{-0.15cm}
\end{equation}

Therefore, according to the definition of the effective capacity, the inequality \eqref{IneqCd} holds.

\vspace{-0.2cm}
\section*{Appendix C \\ Proof of Theorem 3}
\vspace{-0.2cm}

Defining a quadratic Lyapunov function
\vspace{-0.15cm}
\begin{equation} \label{rhoDef}
	\Delta\left(\mathbf{X}_t\right) = \mathbf{X}_t^T\mathbf{P}\mathbf{X}_t
	\vspace{-0.15cm}
\end{equation}
with $\mathbf{P}$ given as a positive definite matrix.

According to \textit{LaSalle's invariance principle} \cite{LyaTheory}, the sufficient conditions of an asymptotic convergent system is 
\vspace{-0.15cm}
\begin{equation} \label{defStability}
	\mathbb{E}\left[\Delta\left(\mathbf{X}_{t+1}\right) \mid \mathbf{X}_t\right] \leq  \Delta\left(\mathbf{X}_t\right).
	\vspace{-0.15cm}
\end{equation}
Substituting \eqref{Var} and \eqref{stateFunc_complex} into \eqref{defStability}, the theorem can be proved.

\begin{acronym}
	\acro{AGV}{automated guided vehicle}
	\acro{MPC}{model predictive control}
\end{acronym}
\vspace{-0.1cm}
\section*{Acknowledgment}
\vspace{-0.1cm}
This work was supported in part by the National Natural Science Foundation of China (NSFC) under Grant 92267202,
in part by the National Key Research and Development Program of China under Grant 2020YFA0711303, and in part by the National Natural Science Foundation of China (NSFC) under Grant 62271081, and U21B2014.

\end{spacing}

%

\end{document}